\documentclass[prb,twocolumn,showkeys,amsmath,amssymb]{revtex4-1}
\usepackage{latexsym}
\usepackage{graphicx}
\usepackage{times}
\usepackage{color}
\usepackage{amsmath}
\usepackage{dcolumn}
\usepackage{amssymb,bm,euscript}
\usepackage{bm}
\usepackage{tabularx}
\usepackage{float}
\usepackage{physics}
\usepackage[colorlinks=true, citecolor=blue, linkcolor=blue, urlcolor=blue, anchorcolor=blue]{hyperref}

\begin{document}

\title{Topological transition in monolayer blue phosphorene with transiton-metal adatom under stain}

\author{Ge Hu and Jun Hu}
\email[]{E-mail: jhu@suda.edu.cn}
\affiliation{School of Physical Science and Technology \& Jiangsu Key Laboratory of Thin Films, Soochow University, Suzhou 215006, China.}

             
\begin{abstract}
We carried out first-principles calculations to investigate the electronic properties of the monolayer blue phosphorene (BlueP) decorated by the group-IVB transition-metal adatoms (Cr, Mo and W), and found that the Cr-decorated BlueP is a magnetic half metal, while the Mo- and W-decorated BlueP are semiconductors with band gaps smaller than 0.2 eV. Compressive strains make the band gaps close and reopen and band inversions occur during this process, which induces topological transitions in the Mo-decorated BlueP (with strain of $-5.75\%$) and W-decorated BlueP (with strain of $-4.25\%$) from normal insulators to topological insulators (TIs). The TI gap is 94 meV for the Mo-decorated BlueP and $150\sim220$ for the W-decorated BlueP. Our findings pave the way to engineer topological phases in the monolayer BlueP with transition-metal adatoms.
\end{abstract}



\maketitle

\section{Introduction}

Advances in modern technologies accelerate miniaturization of electronic devices, which drives the endeavors for exploring exotic materials in nano scale. Ever since the discovery of graphene which was later predicted to be a topological insulator (TI), \cite{graph-0, graph-1, KaneMele1, KaneMele2} two-dimensional (2D) crystals have received extensive research interest because their unique physical and chemical properties that are not found in their bulk counterparts are promising for applications in the future electronics and spintronics devices. So far, various types of 2D materials have been fabricated such as graphene, hexagonal boron nitride, transition metal dichalcogenide monolayers, and so on, with the electronic properties varying from semiconductors, half metals, metals, TIs and so on. \cite{2D-review1, TMD-review, 2D-review2} Among them, the 2D TIs which are a new state of condensed matters inspire great interest in recent years, because they exhibit intriguing quantum spin Hall (QSH) states. Usually, a QSH state is characterized by the combination of insulating bulk state and quantized helical conducting edge state which provides intrinsic spin lock and is robust against elastic backscattering and localization, so that the ITs are ideal for various applications that require dissipationless spin transport. \cite{KaneMele1, KaneMele2, ZhangBernevig1, HgTeTheory, KaneReview, QiReview} Although QSH state was firstly predicted in graphene, it is still difficult to obtain practical 2D TIs with sizable TI gaps to frustrate thermal fluctuations at high temperature.

Recently, the family of 2D elemental phosphorus (termed as phosphorene) has attracted great attention, \cite{phosphorene} since semiconducting few-layer black phosphorene (BlackP) has been fabricated. \cite{blackP-00, blackP-01} The field-effect transistors made of few-layer BlackP exhibit high charge-carrier mobilities up to 1000 cm$^2$V$^{-1}$S$^{-1}$, \cite{blackP-01, blackP-02, blackP-03} which demonstrates the potential for applications in nanoelectronics devices. Apart from the well-known BlackP, other forms of 2D phosphorene such as the monolayer blue phosphorene (BlueP), $\gamma-$phosphorene, and $\delta-$phosphorene have also been reported. \cite{blueP-01, blueP-02} In particular, the monolayer BlueP which crystallizes buckled graphene-like hexagonal structure has been grown on the Au(111) surface. \cite{blueP-03, blueP-04, blueP-05, blueP-06} The band gap ($E_g$) of the free-standing monolayer BlueP was predicted to be $\sim$ 2 eV at the generalized gradient approximation (GGA) of the density functional theory (DFT) level, \cite{blueP-01} while the measurement in experiment for the monolayer BlueP grown on the Au(111) surface yields a smaller band gap of 1.1 eV, partially due to the interaction between the BlueP and metallic substrate. \cite{blueP-04} Nevertheless, the monolayer BlueP maintains its semiconducting character, even though it is grown on a metallic substrate, unlike the case of silicene on Ag(111). \cite{Silicene} Accordingly, the monolayer BlueP is expected to be a promising candidate as a 2D channel material for electronic and optoelectronic devices. \cite{blueP-07}

It is remarkable that a rich diversity of electronic and magnetic properties may be achieved by decorating 2D phosphorene with metal adatoms, especially for the BlackP and BlueP. \cite{Adatom-1, Adatom-2, Adatom-3, Adatom-4, Adatom-5, Adatom-6, Adatom-7} For instance, metallic 2D phosphorene may be obtained with alkali and alkaline-earth adatoms, some transition-metal adatoms such as Cr may results in half-metallicity, and most 3d transition-metal adatoms are magnetic on 2D phosphorene. On the other hand, it was found that a variety of ways could be used to turn monolayer 2D phosphorene from normal insulator into TIs. A monolayer BlackP undergoes topological transition via electric field, doping, hydrostatic pressure, uniaxial strain, or even polarized laser. \cite{Tune-1, Tune-2, Tune-3, Tune-4, Tune-5, Tune-6} A monolayer BlueP may become TI, even through its original electronic band gap is much larger than that of the monolayer BlackP, through oxidization, hydrogenation, fluorination, or strain. \cite{Tune-7, Tune-8, Tune-9} These investigations thus open a route toward the realization of topological phases in a monolayer 2D phosphorene. Clearly, it is interesting whether transition-metal adatoms on a monolayer 2D phosphorene produce topological phases, as that demonstrated in graphene. \cite{Hu-1, Hu-2, Hu-3}

In this paper, we chose the monolayer BlueP as the prototype to explore the possibility of producing TI states in 2D phosphorene with transition-metal adatoms. Based on first-principles calculations, we found that the monolayer BlueP with either Mo adatoms or W adatoms may become TI under certain compressive strain. The TI gap of the Mo-adsorbed BlueP is about 94 meV and that of W-adsorbed BlueP is about $150\sim220$. Analysis of the electronic structures reveals that the TI states originate from the band inversion between different components of the $d$ orbital of the adatoms.

\section{Computational methods}

The structural and electronic properties were calculated with DFT as implemented in the Vienna {\it ab-initio} simulation package. \cite{VASP1, VASP2} The interaction between valence electrons and ionic cores was described within the framework of the projector augmented wave (PAW) method. \cite{PAW1,PAW2} The spin-polarized GGA was used for the exchange-correlation potentials and the spin-orbit coupling (SOC) effect was invoked self-consistently. \cite{PBE} The energy cutoff for the plane wave basis expansion was set to 400 eV. The monolayer BlueP has a buckled graphene-like hexagonal primitive cell with two phosphorous atoms per unit cell, as shown in Fig. \ref{fig1-struct}(a). The optimized lattice constant is 3.275 {\AA}, and the buckling height is 1.24 {\AA}, in agreement with previous experimental measurement. \cite{blueP-04} In this geometry, the  P$-$P bond length and P$-$P$-$P bond angle is 2.26{\AA} and 92.92$^{\circ}$, respectively. To explore the electronic properties of transition-metal adatoms on the monolayer BlueP, we used a 2$\times$2 supercell and the 2D Brillouin zone was sampled by a 15$\times$15 k-grid mesh. There are four high symmetric adsorption sites: the top ($T$) and valley ($V$) sites over the top and bottom P atoms, respectively; the hollow site ($H$) above the two P layers; the center site ($C$) between the two P layers, as notated in Fig. \ref{fig1-struct}(a). With an adatom at each of these sites, a vacuum of 15 {\AA} is added to avoid fake interaction between neighboring monolayers which is produced by the periodic boundary condition. Both the lattice constant and atomic positions were fully relaxed with a criterion that requires the forces on each atom smaller than 0.01 eV/{\AA}. The band topology is characterized by the topological invariant $\mathbb{Z}_2$, with $\mathbb{Z}_2=1$ for TIs and $\mathbb{Z}_2=0$ for normal insulators. \cite{Z2}  We adopted the so-called $n-$field scheme to calculate $\mathbb{Z}_2$. \cite{n-field-0, n-field-1, n-field-2}

\section{Results and discussion}

\begin{table}
 \centering
 \caption{Binding energy (in eV) for an adatom at different adsorption sites: center ($C$), hollow ($H$), valley ($V$) and top ($T$).}
 \tabcolsep0.21in             
 \begin{tabular}{cccccccccc}
   \hline
   \hline
Aadatom & $C$ & $H$ & $V$ & $T$  \\
   \hline
Cr & 0.09 & 1.24 & 1.48 & 0.75 \\
Mo & 1.84 & 1.71 & 2.61 & 0.94 \\
W  & 3.10 & 2.84 & 3.38 & 1.41 \\
   \hline
   \hline
 \end{tabular}
\end{table}

\begin{figure}[t]
\centering
\includegraphics[width=8.5cm]{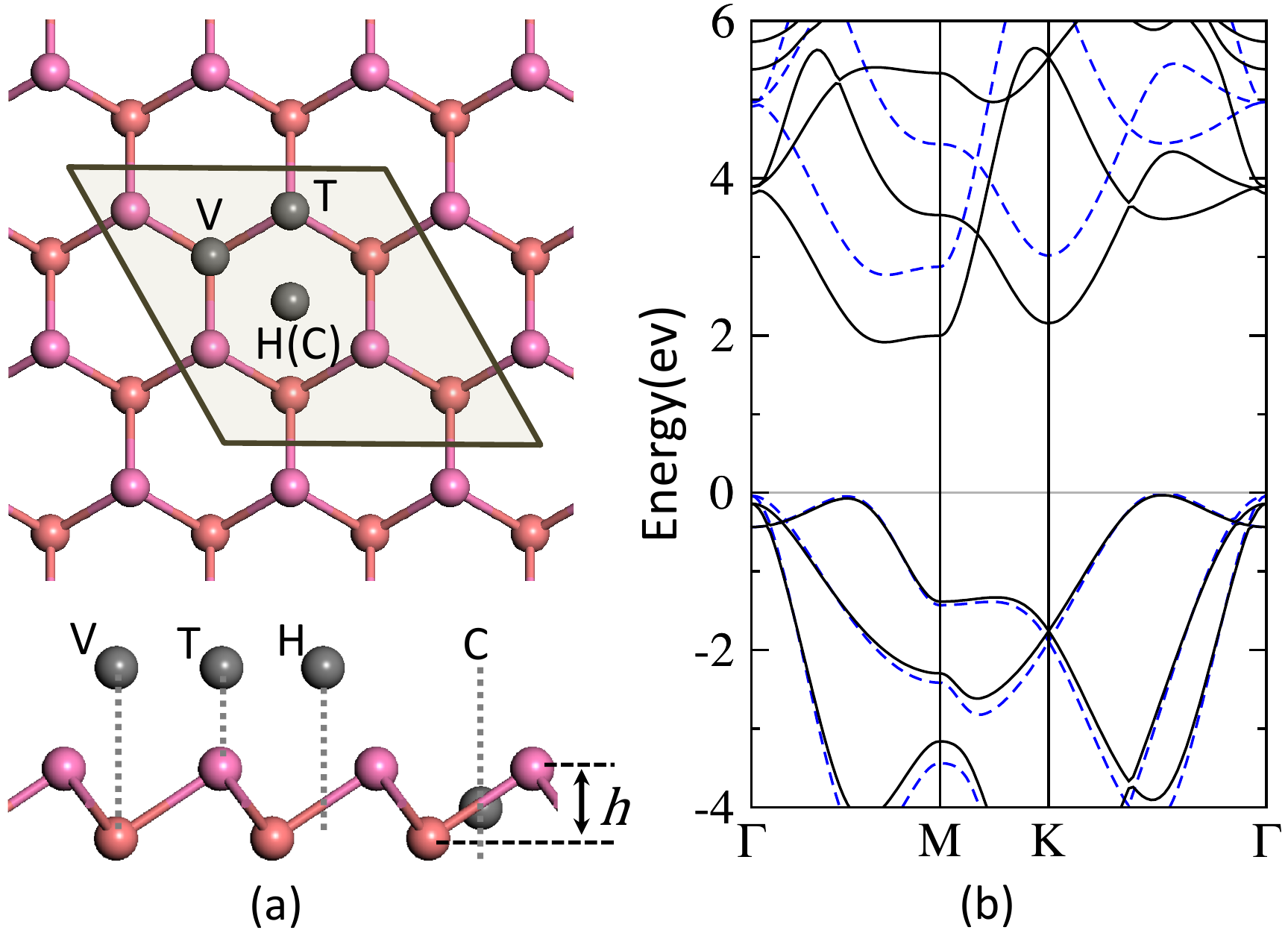}
\caption{(a) Top (upper panel) and side (lower panel) views of a monolayer BlueP. The coral and plum spheres stand for the two sublayers of P atoms. The dark spheres stand for adatoms at different adsorption sites: hollow ($H$), center($C$), valley($V$), and top ($T$). Note that the hollow site is above the BlueP plane, while the center site is at the middle point between the two sublayers. The parallelogram indicates a 2$\times $2 supercell. (b) PBE (black solid curves) and HSE06 (blue dashed curves) band structure without SOC of pure blue phosphorene. The VBM is set to 0 eV.
}\label{fig1-struct}
\end{figure}

\begin{figure}[t]
\centering
\includegraphics[width=8.5cm]{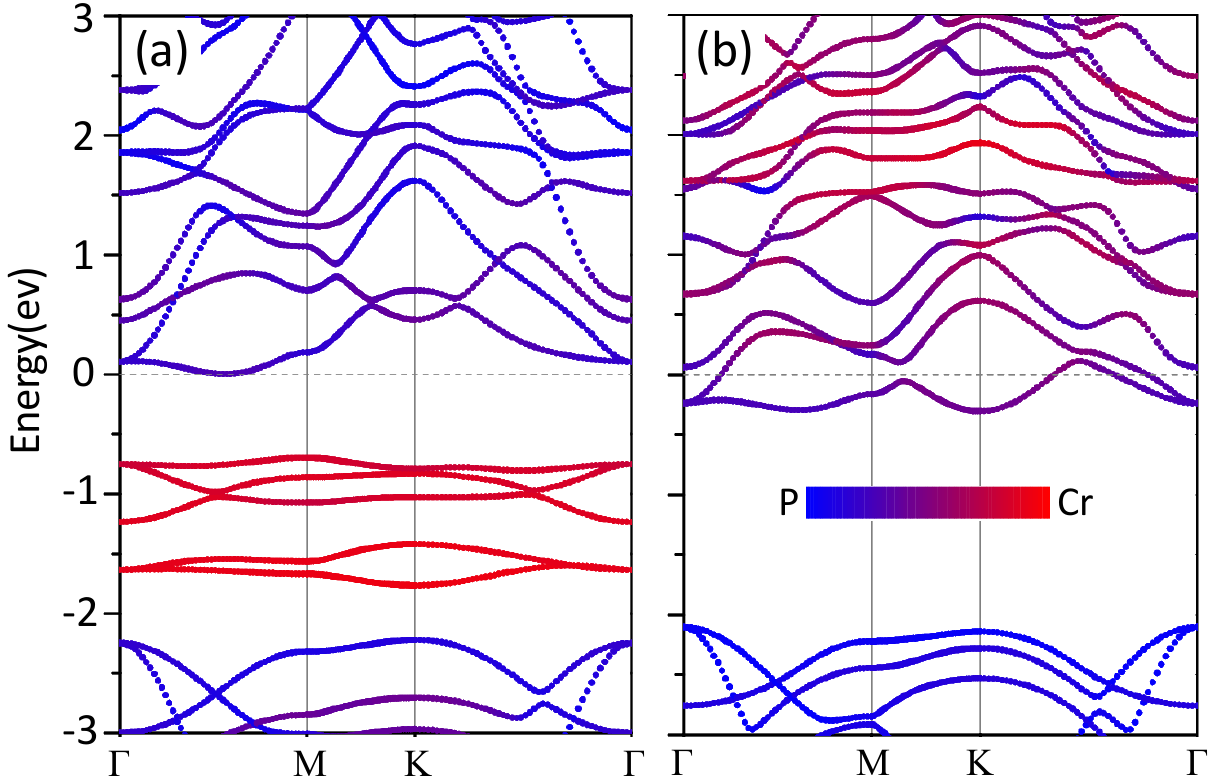}
\caption{(a) Spin-up and (b) spin-down band structures of the monolayer BlueP with a Cr adatom adsorbed at a valley site. The color bar indicates the weights from the host P atoms and Cr adatom. The Fermi energy is set to 0 eV.
}\label{fig2-Cr}
\end{figure}

\begin{figure*}[t]
\centering
\includegraphics[width=14cm]{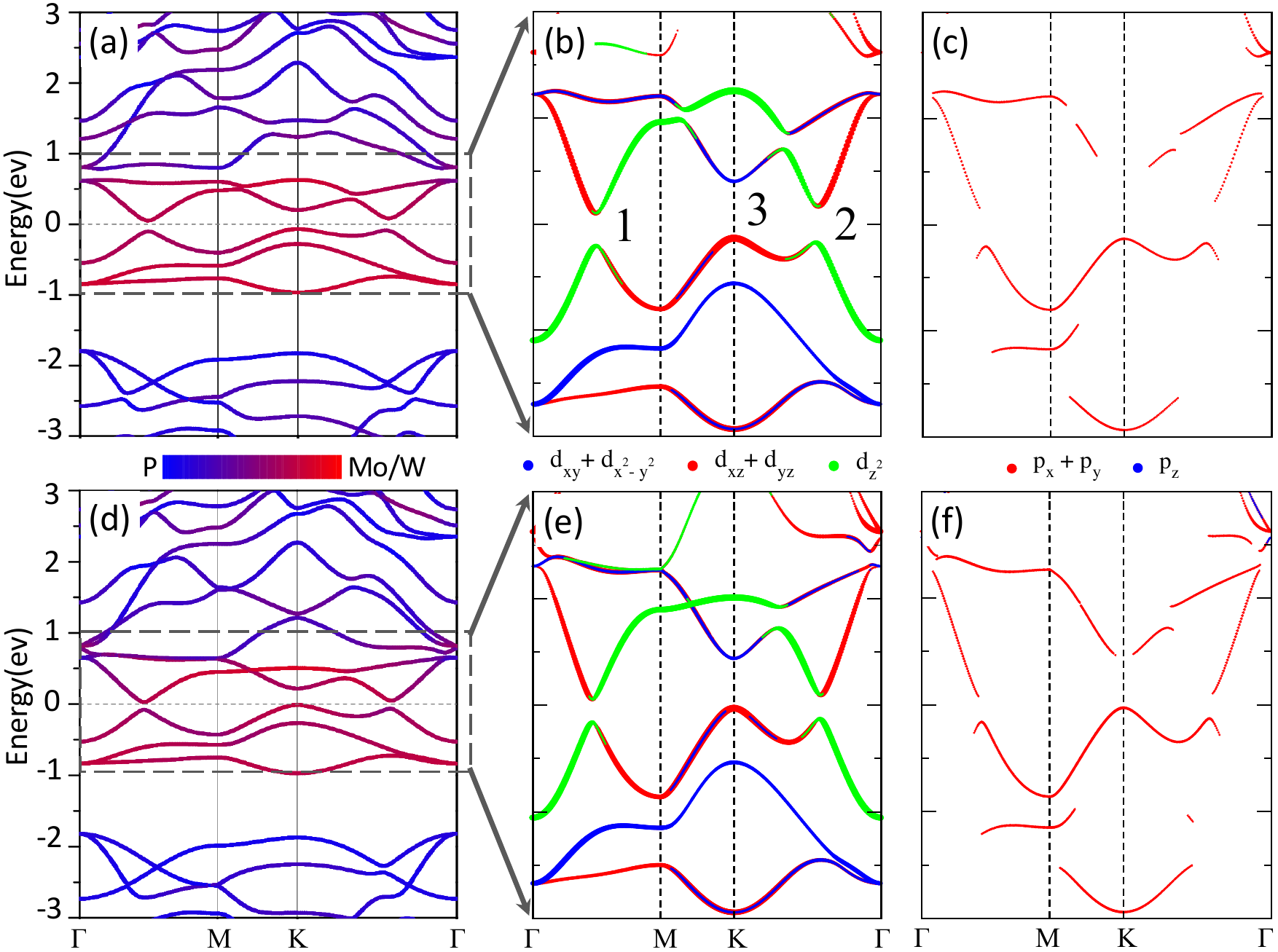}
\caption{Band structures of the monolayer BlueP with a (a--c) Mo or (d--f) W adatom adsorbed at a valley site. The Fermi energy is set to 0 eV. (a) and (d) The complete band structures with the weights of the host P atoms and Mo or W adatom indicated by the color bar. (b) and (e) Bands between $-1.0$ eV and 1.0 eV with the weights of the $d$ orbitals of the Mo or W adatom. (c) and (f) Bands between $-1.0$ eV and 1.0 eV with the weights of the $p$ orbitals of the host P atom underneath the adatom.
}\label{fig3-Mo-W}
\end{figure*} 
  
We first calculated the band structure of the monolayer BlueP, as shown in Fig. \ref{fig1-struct}(b). It can be seen that the monolayer BlueP is an indirect-band-gap semicondutor with the valence band maximum (VBM) and the conduction band minimum (CBM) near the midpoint along the path {$\overline{\bf\Gamma{K}}$} and {$\overline{\bf\Gamma{M}}$}, respectively. The band gap at the GGA level is 2 eV, in accordance with the previous calculations. \cite{blueP-01} It is known that the regular GGA calculations usually underestimate band gaps of semiconductors, so we carried out further calculation with the hybrid functional HSE06 \cite{HSE06} to obtain more accurate band gap. As seen from Fig. \ref{fig1-struct}(b), the dispersions of the bands from the HSE06 calculation are similar to those from the GGA calculations, but the band gap increases significantly to 2.8 eV, which indicates that the monolayer BlueP is a wide-band-gap semiconductor.

The group-VIB transition-metal elements (Cr, Mo and W) were chosen to be the adatoms. To find the most stable adsorption configure, we calculated the binding energy ($E_b$) with one adatom at each of the four different adsorption sites as indicated in Fig. \ref{fig1-struct}(a) in a 2$\times $2 supercell. $E_b$ is calculated as
\begin{equation}
 E_b=E(BlueP)+E(Adatom)-E(Adatom/BlueP), 
\end{equation}
where $E(BlueP)$, $E(Adatom)$ and $E(Adatom/BlueP)$ are the total energies of the pure monolayer BlueP, isolated transition-metal atom and the BlueP with adatom, respectively. As listed in Table I, all the adatoms prefer the $V$ site, in line with the previous calculations. \cite{Adatom-1, Adatom-3, Adatom-5} In this configuration, the adatom mainly bonds to the P atom underneath and the bonding between the adatom and the three neighboring P atoms which form the upper bound of the valley is weaker. For all adatoms, the amplitudes of $E_b$ at the $V$ site are large, which implies strong hybridization between the BlueP and adatoms. In addtion, $E_b$ increases from $3d$ to $5d$ adatom, because of the increasing chemical activity for larger adatom. Moreover, the Cr adatom at the $V$ site is magnetic, with magntic moment of 4.1 $\mu_B$, while both Mo and W adatoms are nonmagnetic.

Figure \ref{fig2-Cr} shows the element-resolved band structure for the monolayer BlueP with a Cr adatom at the $V$ site.  Attractively, the system is half-metallic, the same as the previous reports. \cite{Adatom-1, Adatom-3, Adatom-5} The bands in the energy range of $-2.5\sim-2.0$ eV are comprised of pure P states. In the majority spin channel [Fig. \ref{fig2-Cr}(a)], a few bands (from $-1.7$ to $-0.7$ eV) appear in the band gap of the BlueP ($-2.2\sim0$ eV), and they are contributed almost completely from the $3d$ orbitals of the Cr adatom. In the minority spin channel [Fig. \ref{fig2-Cr}(a)], two bands cross the Fermi level, and they are from the hybridization between the P and Cr atoms as indicated by the color bar. Clearly, there is a pseudo band gap in the energy range of $-2.1\sim-0.3$ eV, corresponding to the original band gap of the BlueP. Therefore, the Cr adatom donates electron charge to the BlueP, which pushes the Fermi level to the conduction band of the BlueP. Along with the hybridization between the BlueP and Cr adatom, the whole system becomes a magnetic half metal which may be used in nano-spintronics devices.

The monolayer BlueP with either Mo or W adatom is nonmagnetic and semiconducting. However, the band gap decreases strikingly down to 0.16 eV for the Mo-adsorbed BlueP and 0.11 for the W-adsorbed BlueP, as shown in Fig. \ref{fig3-Mo-W}(a) and \ref{fig3-Mo-W}(d). In both cases, the bands near $-2.0$ eV are contributed from the P atoms, similar to the case with the Cr adatom. The bands around the Fermi level ($-1.0\sim1.0$ eV) mainly originate from the adatoms, slightly hybridized with the P atom, as indicated by the color bar. To further identify the atomic characteristics of these bands in details, we calculated the weights of each atomic orbital for all the bands, as plotted in Fig. \ref{fig3-Mo-W}. Due to the local symmetry, the $d$ orbitals of the adatoms split into three groups: in-plane orbitals with degenerate $d_{xy}$ and $d_{x^2-y^2}$, cross orbitals with degenerate $d_{xz}$ and $d_{yz}$, and perpendicular orbital with $d_{z^2}$. From Fig. \ref{fig3-Mo-W}, it can be seen that the hybridization in the energy range of $-1.0\sim1.0$ eV mainly occurs between $d_{xz/yz}$ of the adatom and $p_{x/y}$ of the P atom underneath, while the other orbitals do not hybridize notably. Moreover, the three neighboring P atoms surrounding the adatom do not contribute to these bands. Interestingly, there are obvious intra-atomic hybridizations between the $d$ orbitals of the adatoms, which produce three local band gaps around the Fermi level. The two gaps near the midpoints along the paths {$\overline{\bf\Gamma{M}}$} and {$\overline{\bf\Gamma{K}}$} (denoted as gap-1 and gap-2) derive from the hybridization between $d_{xz/yz}$ and $d_{z^2}$, while the gap at {\bf{K}} point (denoted as gap-3) is mainly from $d_{xy/x^2-y^2}$ and $d_{xz/yz}$. This feature offers the opportunity to engineer topological phases, because band inversion could be induced by biaxial strain for small-band-gap semiconductors. \cite{band-inversion-1, band-inversion-2} By applying tensile or compressive strain, the small gap may close at certain strength of strain and then reopen as the strain further increases. This closing-reopening process results in band inversion beside the gap, which is a typical characteristic for TIs.

\begin{figure}[t]
\centering
\includegraphics[width=8cm]{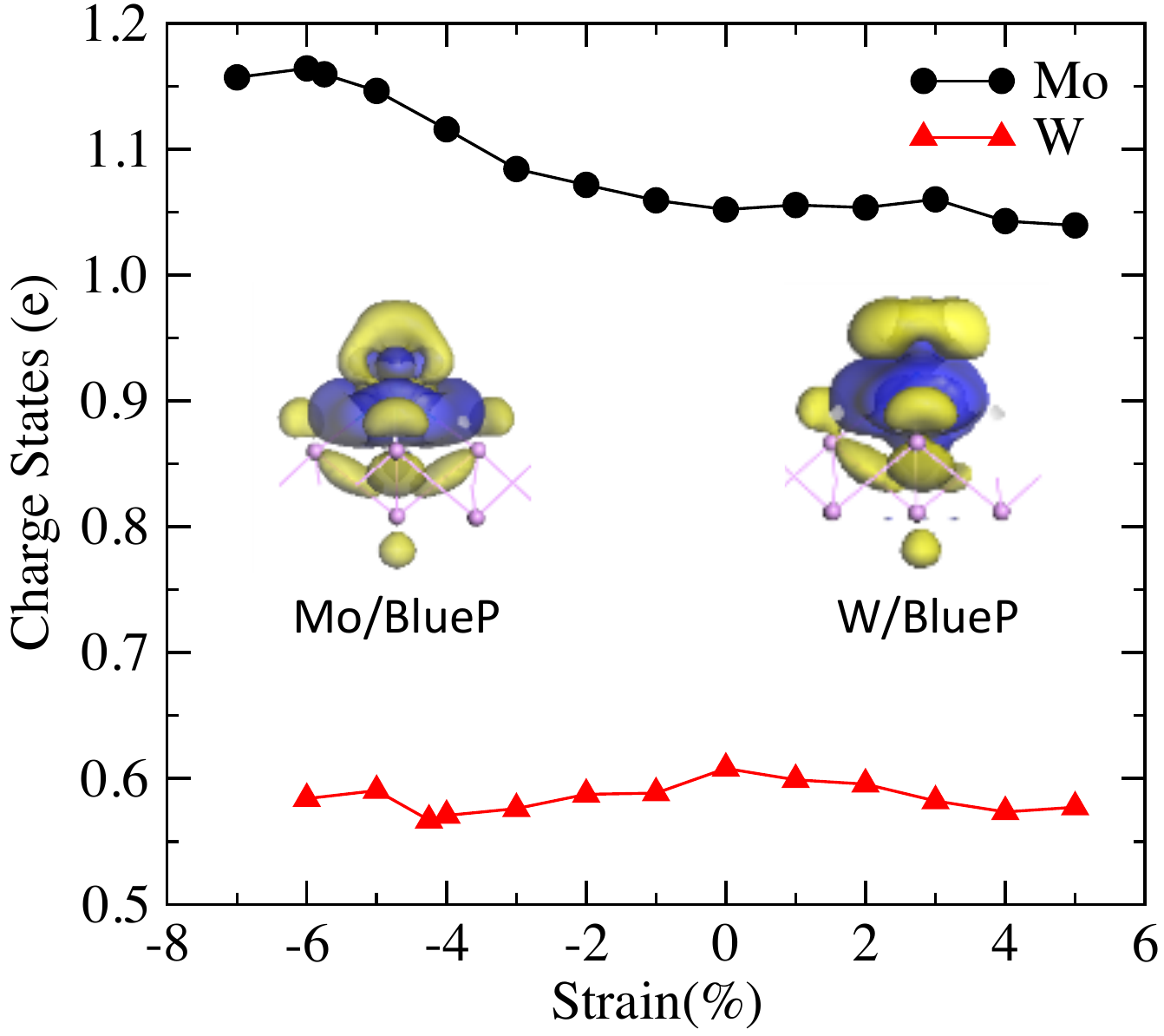}
\caption{Charge states of the Mo and W adatoms at a valley site under different biaxial strains. The insets shows the electron charge redistribution caused by the interaction between the host P atoms and adatoms. The yellow and blue regions indicate electron charge gain and loss, respectively.
}\label{fig4-charge}
\end{figure}

\begin{figure}[t]
\centering
\includegraphics[width=8.5cm]{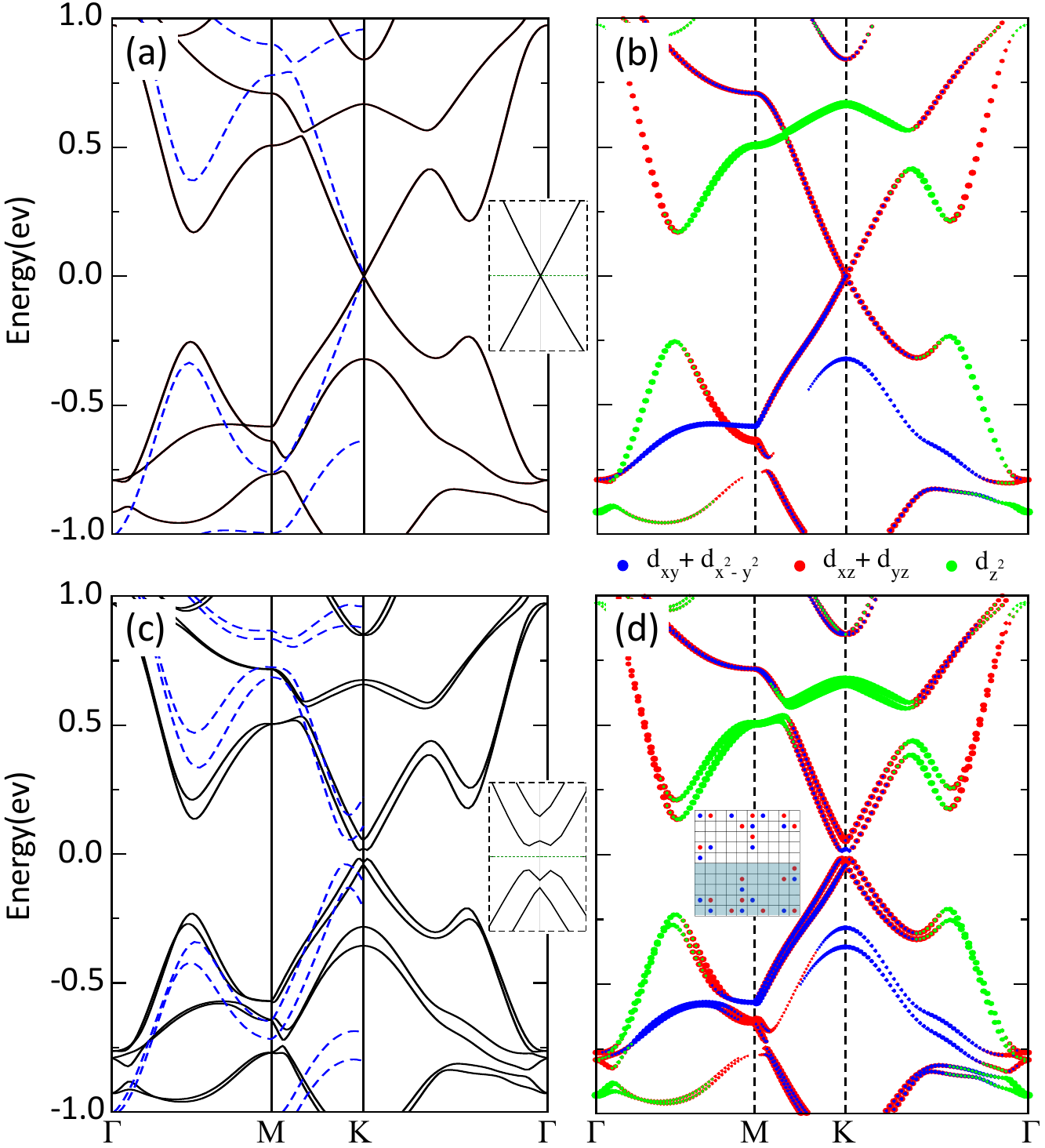}
\caption{Band structures of the monolayer BlueP with a Mo adatom adsorbed at a valley site under compressive strain of $-5.75\%$ (a,b) without and (c,d) with the spin-orbit coupling. The Fermi energy is set to 0 eV. The blue dashed curves in (a) and (c) are bands calculated by the HSE06 method. The insets (a) and (c) show low-energy bands near the Fermi energy. (b) and (d) Bands with the weights of the $d$ orbitals of the Mo adatom. The inset in (d) shows the $n-$field configuration. The nonzero points are denoted by red ($n=1$) and blue ($n=-1$) dots, respectively. The $\mathbb{Z}_2$ invariant is obtained by summing all the $n$ in half Brillouin zone marked by the shadow.
}\label{fig5-strain-Mo}
\end{figure}

To investigate the effect of biaxial strain on the electronic property, we applied biaxial strain ($\varepsilon$) from $-7$\% to 5\% where negative and positive values denote compressive and tensile strains, respectively. Intuitively, the strain might alter the hybridization between the adatom and BlueP, which ought to cause change of charge state of the adatoms. As shown in Fig. \ref{fig4-charge}, compressive strain make the charge state of the Mo adatom increase from $+1.05e$ at zero strain to maximum value of $+1.17e$ at strain of $-6\%$. On the contrary, tensile strain does not leads visible change of the charge states of the Mo adatom up to strain of $5\%$. For the W adatom, the overall charge states ($+0.57e\sim+0.61e$) are much smaller than those of the Mo adatom ($+1.04e\sim+1.17e$). In addition, both compressive and tensile strains reduce the charge state of the W adatom slightly. The insets in Fig. \ref{fig4-charge} show the details of the charge redistribution caused by the hybridization between the adatom and BlueP. It can be seen that the charge transfer happens not only between the adatom and its neighboring P atoms but also between different orbitals of the adatom. Nevertheless, the charge transfer from the adatoms to the BlueP dominates, which results in positive charge state of the adatoms.

The band structures with strains are plotted in Fig. S1 and S2. For both Mo and W adatoms, gap-3 undergoes closing-reopening process under compressive strain, while this happens for gap-1 and gap-2 under tensile strain. Since gap-3 locates at highly symmetric {\bf K} point, it may be more interesting than the other two gaps. Hence, we focus on gap-3 in the following. To get more accurate critical point of the closing-reopening process, we set finer mesh of strain around the critical point. We found that gap-3 in the Mo-adsorbed BlueP closes at compressive strain of $-5.75\%$, and the corresponding strain is $-4.25\%$ for the W-adsorbed BlueP. Figure \ref{fig5-strain-Mo}(a) shows that two bands cross the Fermi level at {\bf K} point and they are linearly dispersive near the Fermi level. This attribute implies that the low-energy electrons behave as massless Dirac Fermions as in graphene. \cite{KaneMele1, KaneMele2} Moreover, these bands are mainly from the $d_{xy/x^2-y^2}$ and $d_{xz/yz}$ orbitals of Mo, mixed slightly with the $p_{x/y}$ orbital of the P atom underneath, as presented in Fig. \ref{fig5-strain-Mo}(b) and \ref{fig5-strain-Mo}(c). Including the SOC effect, the degeneracy of the linear bands is lifted due to the Rashba effect and a gap of 32.5 meV opens at {\bf K} point [Fig. \ref{fig5-strain-Mo}(d)]. The atomic-orbital resolved bands in Fig. \ref{fig5-strain-Mo}(d) shows that the highest occupied and lowest unoccupied levels at {\bf K} point are contributed completely from $d_{xz/yz}$ and $d_{xy/x^2-y^2}$, respectively, which indicates band inversion induced by the SOC effect. To confirm that the band inversion indeed results in topological phase in the Mo-adsorbed BlueP, we calculated $\mathbb{Z}_2$ with the $n-$field method. \cite{n-field-0, n-field-1, n-field-2} By summing the positive and negative $n-$field numbers over half of the torus as indicated in the inset in Fig. \ref{fig5-strain-Mo}(d), we obtained $\mathbb{Z}_2=1$ for the Mo-adsorbed BlueP, clearly demonstrating the nontrivial band topology.

As mentioned above, the GGA calculations usually underestimate band gaps of semiconductors. Therefore, it is necessary to verify the Dirac states in the Mo-adsorbed BlueP at the HSE06 level. To save computational resources, we only calculated the band structure of the path ${\bf \Gamma - M - K}$ which is enough to capture the feature of the bands near the Fermi level. Interestingly, the linearly dispersive Dirac bands maintain in the HSE06 calculations, as seen in Fig. \ref{fig5-strain-Mo}(a). The SOC effect induces a gap of 94.3 eV near {\bf K} point [Fig. \ref{fig5-strain-Mo}(c)], about three times of gap-3 at the GGA level. Furthermore, the topological invariant $\mathbb{Z}_2$ is still 1, which demonstrates that the Mo-adsorbed BlueP is a TI at the HSE06 level.

\begin{figure}
\centering
\includegraphics[width=8.5cm]{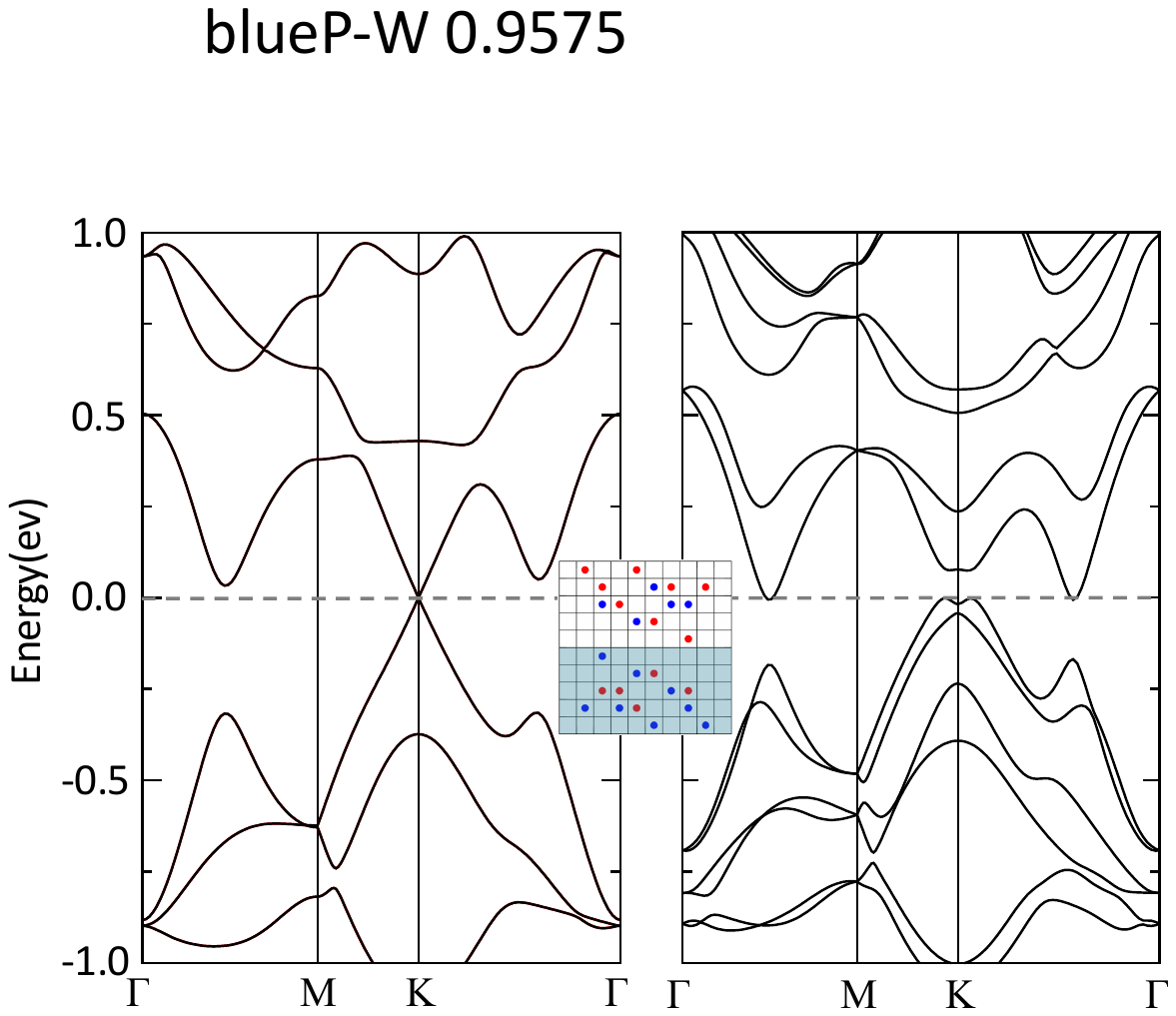}
\caption{Band structures of the monolayer BlueP with a W adatom adsorbed at valley site under compressive strain of $-4.25\%$ (a) without and (b) with the spin-orbit coupling. The Fermi energy is set to 0 eV. The inset shows the $n-$field configuration.
}\label{fig6-strain-W}
\end{figure}

The band structures of the W-adsorbed BlueP are plotted in Fig. \ref{fig6-strain-W}. Similar to the Mo-adsorbed BlueP, the two bands crossing the Fermi level disperse linearly, and the SOC effect leads to a gap of 76.1 meV near {\bf K} point. It should be pointed out that the SOC effect also makes the upper bands of gap-1 and gap-2 decrease down to the Fermi level, so it seems that the W-adsorbed BlueP is a semimetal. However, the W-adsorbed BlueP is actually a TI because the topological invariant $\mathbb{Z}_2$ is 1 as calculated with the $n-$field method [see the inset in Fig. \ref{fig6-strain-W}]. Although we did not performed HSE06 calculations for the W-adsorbed BlueP, it might be expected that the amplitude of the nontrivial topological gap is $2\sim3$ times of that at the GGA level, i.e., the TI gap could be as large as $150\sim220$ meV, which is promising for application at high temperature.

\section{Conclusion}

Based on first-principles calculations, we studied the electronic properties of the monolayer BlueP with group-IVB transition-metal adatoms (Cr, Mo and W). We found that the Cr-adsorbed BlueP is magnetic half metal, while the Mo- and W-adsorbed BlueP are semiconductors with band gap smaller than 0.2 eV. The hybridization between the host P atoms and the adatom generates a few bands in the original band gap of the monolayer BlueP, leading the reduction of the band gaps of the Mo- and W-adsorbed BlueP. Intriguingly, compressive strains of $-5.75\%$ and $-4.25\%$ turn the Mo- and W-adsorbed BlueP into TI, respectively. The TI gap is $\sim$94 meV for the Mo-adsorbed BlueP and $150\sim220$ for the W-adsorbed BlueP. The band inversion between different components of the $d$ orbital of the adatoms is responsible for the topological transition. Our findings manifest that the Mo- and W-adsorbed BlueP may be good platforms to investigate the QSH effect in 2D TIs and promising for applications in spintronics devices.

\section*{Acknowledgements}

This work is supported by the National Natural Science Foundation of China (11574223), the Natural Science Foundation of Jiangsu Province (BK20150303).



\begin{thebibliography}{99}
\bibitem{graph-0} K. S. Novoselov, A. K. Geim, S. V. Morozov, D. Jiang, Y. Zhang, S. V. Dubonos, I. V. Grigorieva, A. A. Firsov, Science, 2004, {\bf 306}, 666--669.

\bibitem{graph-1} K. S. Novoselov, D. Jiang, F. Schedin, T. J. Booth, V. V. Khotkevich, S. V. Morozov, and A. K. Geim, PNAS, 2005, {\bf 102}, 10451--10453.


\bibitem{KaneMele1} C. L. Kane and E. J. Mele, Phys. Rev. Lett., 2005, {\bf 95}, 146802.

\bibitem{KaneMele2} C. L. Kane and E. J. Mele, Phys. Rev. Lett., 2005, {\bf 95}, 226801.

\bibitem{2D-review1}M. Xu, T. Liang, M. Shi, and H. Chen, Chem. Rev., 2013, {\bf 113}, 3766--3798.

\bibitem{TMD-review} M. Chhowalla, H. S. Shin, G. Eda, L.-J. Li, K. P. Loh, and H. Zhang, Nat. Chem., 2013, {\bf 5}, 263--275.

\bibitem{2D-review2} Z. Wang, J. Qiu, X. Wang, Z. Zhang, Y. Chen, X. Huang, and W. Huang, Chem. Soc. Rev., 2018, {\bf 47}, 6128--6174. 

\bibitem{ZhangBernevig1}  B. A. Bernevig and S. C. Zhang, Phys. Rev. Lett., 2006, {\bf 96}, 106802.

\bibitem{HgTeTheory} B. A. Bernevig, T. L. Hughes and S. C. Zhang, Science, 2006, {\bf 314}, 1757--1761.


\bibitem{KaneReview} M. Z. Hasan and C. L. Kane, Rev. Mod. Phys., 2010, {\bf 82}, 3045--3067.

\bibitem{QiReview} X. L. Qi and S. C. Zhang, Rev. Mod. Phys., 2011, {\bf 83}, 1057--1110.

\bibitem{phosphorene} E. S. Reich, Nature, 2014, {\bf 506}, 19.

\bibitem{blackP-00} H. Liu, A. T. Neal, Z. Zhu, D. Tomanek, and P. D. Ye, ACS Nano, 2014, {\bf 8}, 4033--4041.

\bibitem{blackP-01} L. Li, Y. Yu, G. J. Ye, Q. Ge, X. Ou, H. Wu, D. Feng, X. H. Chen, and Y. Zhang, Nat. Nanotechnol., 2014, {\bf 9}, 372.

\bibitem{blackP-02} M. Buscema, D. J. GroenendijkSofya, I. Blanter, G. A. Steele, H. S. J. van der Zant, and A. Castellanos-Gomez, Nano Lett., 2014, {\bf 14}, 6, 3347--3352.

\bibitem{blackP-03} F. N. Xia, H. Wang, and Y. C. Jia, Nat. Commun., 2014, {\bf 5}, 4458.

\bibitem{blueP-01}Z. Zhu and D. Tom\'anek, Phys. Rev. Lett., 2014, {\bf 112}, 176802.

\bibitem{blueP-02} J. Guan, Z. Zhu, and D. Tom\'anek, Phys. Rev. Lett., 2014, {\bf 113}, 046804.
 
\bibitem{blueP-03} J. Zeng, P. Cui, and Z. Zhang, Phys. Rev. Lett., 2017, {\bf 118}, 046101.
 
\bibitem{blueP-04} J. L. Zhang, S. Zhao, C. Han, Z. Wang, S. Zhong, S. Sun, R. Guo, X. Zhou, C. D. Gu, K. D. Yuan, Z. Li, and W. Chen, Nano Lett., 2016, {\bf 16}, 4903-4908.

\bibitem{blueP-05} C. Gu, S. Zhao, J. L. Zhang, S. Sun, K. Yuan, Z. Hu, C. Han, Z. Ma, L. Wang, F. Huo, W. Huang, Z. Li, and W. Chen, ACS Nano, 2017, {\bf 11}, 4943--4949.

\bibitem{blueP-06} N. Han, N. Gao, and J. Zhao, J. Phys. Chem. C, 2017, {\bf 121}, 17893--17899.

\bibitem{Silicene} S. K. Mahatha, P. Moras, V. Bellini, P. M. Sheverdyaeva, C. Struzzi, L. Petaccia, and C. Carbone, Phys. Rev. B, 2014, {\bf 89}, 201416(R).

\bibitem{blueP-07} J. Li, X. Sun, C. Xu, X. Zhang, Y. Pan, M. Ye, Z. Song, R. Quhe, Y. Wang, H. Zhang, Y. Guo, J. Yang, F. Pan, and J. Lu, Nano Res., 2018, {\bf 11}, 1834--1849.

\bibitem{Adatom-1} V. V. Kulish, O. I. Malyi, C. Perssoncd, and P. Wu, Phys. Chem. Chem. Phys., 2015, {\bf 17}, 992--1000.

\bibitem{Adatom-2} T. Hu and J. Hong, J. Phys. Chem. C, 2015, {\bf 119}, 8199--8207.

\bibitem{Adatom-3} X. Sui, C. Si, B. Shao, X. Zou, J. Wu, B.-L. Gu, and W. Duan, J. Phys. Chem. C, 2015, {\bf 119}, 10059--10063.

\bibitem{Adatom-4} Y. Ding and Y. Wang, J. Phys. Chem. C, 2015, {\bf 119}, 10610--10622.

\bibitem{Adatom-5} I. Khan, J. Son, and J. Hong, Phys. Lett. A, 2018, {\bf 382}, 205--209.

\bibitem{Adatom-6} Y. Luo, C. Ren, S. Wang, S. Li, P. Zhang, J. Yu, M. Sun, Z. Sun, and W. Tang,  Nanoscale Res. Lett., 2018, {\bf 13}, 282.

\bibitem{Adatom-7} X.-Q. Lu, C.-K. Wang, and X.-X. Fu, Phys. Chem. Chem. Phys., 2019, {\bf 21}, 11755--11763.

\bibitem{Tune-1} Q. Liu, X. Zhang, L. B. Abdalla, A. Fazzio, and A. Zunger, Nano Lett., 2015, {\bf 15}, 1222--1228.

\bibitem{Tune-2} J. Kim, S. S. Baik, S. H. Ryu, Y. Sohn, S. Park, B.-G. Park, J. Denlinger, Y. Yi, H. J. Choi, and K. S. Kim, Science, 2015, {\bf 349}, 723--726.

\bibitem{Tune-3} S. S. Baik, K. S. Kim, Y. Yi, and H. J. Choi, Nano Lett., 2015, {\bf 15}, 7788--7793.

\bibitem{Tune-4} R. Fei, V. Tran, and L. Yang, Phys. Rev. B, 2015, {\bf 91}, 195319.

\bibitem{Tune-5} E. T. Sisakht, F. Fazileh, M. H. Zare, M. Zarenia, and F. M. Peeters, Phys. Rev. B, 2016, {\bf 94}, 085417.

\bibitem{Tune-6} C. Dutreix, E. A. Stepanov, and M. I. Katsnelson, Phys. Rev. B, 2016, {\bf 93}, 241404.

\bibitem{Tune-7} L. Zhu, S.-S. Wang, S. Guan, Y. Liu, T. Zhang, G. Chen, and S. A. Yang, Nano Lett., 2016, {\bf 16}, 6548--6554.

\bibitem{Tune-8} M. Sun, S. Wang, J. Yu, and W. Tang, Appl. Surf. Sci., 2017,  {\bf 392}, 46--50.

\bibitem{Tune-9} G. Yang, Z. Xu, Z. Liu, S. Jin, H. Zhang, and Z. Ding, J. Phys. Chem. C, 2017, {\bf 121}, 12945--12952.

\bibitem{Hu-1} C. Weeks, J. Hu, J. Alicea, M. Franz and R. Q. Wu, Phys. Rev. X, 2011, {\bf 1}, 021001.

\bibitem{Hu-2} J. Hu, J. Alicea, R. Q. Wu and M. Franz, Phys. Rev. Lett., 2012, {\bf 109}, 266801.

\bibitem{Hu-3} J. Hu, Z. Y. Zhu and R. Q. Wu, Nano Lett., 2015, {\bf 15}, 2074--2078.

\bibitem{VASP1} G. Kresse and J. Furthm\"{u}ller, Comput. Mater. Sci., 1996, {\bf 6}, 15--50.

\bibitem{VASP2} G. Kresse and J. Furthm\"{u}ller, Phys. Rev. B, 1996, {\bf 54}, 11169--11186.

\bibitem{PAW1} P. E. Bl\"{o}chl, Phys. Rev. B, 1994, {\bf 50}, 17953--17979.

\bibitem{PAW2} G. Kresse and D. Joubert, Phys. Rev. B, 1999, {\bf 59}, 1758--1775.

\bibitem{PBE} J. P. Perdew, K. Burke and M. Ernzerho, Phys. Rev. Lett., 1996, {\bf 77}, 3865--3868.

\bibitem{Z2} L. Fu and C. L. Kane, Phys. Rev. B, 2006, {\bf 74}, 195312.

\bibitem{n-field-0} T. Fukui and Y. Hatsugai, J. Phys. Soc. Jpn., 2007, \textbf{76}, 053702.

\bibitem{n-field-1} D. Xiao, Y. Yao, W. Feng, J. Wen, W. Zhu, X. Q. Chen, G. M. Stocks and Z. Zhang, Phys. Rev. Lett., 2010, {\bf 105}, 096404.

\bibitem{n-field-2} W. Feng, J. Wen, J. Zhou, D. Xiao and Y. Yao, Comput. Phys. Commun., 2012, {\bf 183}, 1849--1859.

\bibitem{HSE06} J. Paier, M. Marsman, K. Hummer, G. Kresse, I. C. Gerber, and J. G. \'Angy\'an, J. Chem. Phys., 2006, {\bf 124}, 154709.

\bibitem{band-inversion-1} Z. Zhu, Y. Chen, and U. Schwingenschl\"ogl, Phys. Rev. Lett., 2012, {\bf 108}, 266805.

\bibitem{band-inversion-2} Z. Zhu, Y. Chen, and U. Schwingenschl\"ogl, Phys. Rev. Lett., 2012, {\bf 85}, 235401.

\end{thebibliography}
\end{document}